\documentclass[twocolumns,traditabstract]{aa}
\usepackage{graphicx}
\usepackage{txfonts}
\def\co{CO(1-0)}

\begin{document}
   \title{\object{High resolution mapping of CO(1-0) in NGC6240}
   \thanks{
Based on observations carried out with the IRAM Plateau de Bure Interferometer. IRAM is supported by INSU/CNRS (France), MPG (Germany) and IGN (Spain).
}
}

   \author{C. Feruglio \inst{1}
          \and
	F. Fiore  \inst{2} 
        \and
            E. Piconcelli  \inst{2}
	\and 
        C. Cicone \inst{3}
	\and
          R. Maiolino \inst{3}
                  \and
        R. Davies \inst{4}
	\and
         E. Sturm \inst{4}
          }

   \institute{IRAM - Institut de RadioAstronomie Millim\'etrique 
300 rue de la Piscine, Domaine Universitaire 
38406 Saint Martin d'H\'eres, France, \email{feruglio@iram.fr}
         \and
            INAF- Osservatorio astronomico di Roma, via Frascati 33, 00040 Monteporzio Catone, Italy
            \and
            Cavendish Laboratory, University of Cambridge, 19 J. J. Thomson Ave., Cambridge CB3 0HE, UK
            \and
            Max-Planck-Institut fur Extraterrestrische Physik (MPE), Giessenbachstr. 1, 85748 Garching, Germany
             }


 
  \abstract{
We present high spatial resolution CO(1-0) mapping of the luminous infrared galaxy NGC6240 obtained with the IRAM - Plateau de Bure Interferometer. 
This source is a well-known early-stage merging system hosting two AGN.
We find a broad CO(1-0) line profile, with maximum velocity 800 $\rm km~s^{-1}$ and $\rm FWZI=1400$ $\rm km~s^{-1}$, and displaying several kinematic components, revealing the complexity of the gas dynamics in this system. 
We detect a blueshifted CO emission with velocity in the range -200 and -500 $\rm km~s^{-1}$, peaked around the southern AGN, at the same position where the $\rm H_2$ outflow is located, and with a mass loss rate of $\sim 500~ \rm M_\odot yr^{-1}$. We interpret this blueshifted component as a outflow, originating from the southern nucleus. 
The spatial and spectral match strongly suggests that the CO outflow is connected to the $\rm H_2$ superwind located around the southern nucleus, and to the large scale CO outflow, with similar velocities, extended on scales of 10 kpc. 
The large mass loading factor ($\rm \dot M/SFR \sim 10 $) of the molecular gas suggests that the outflow is likely driven by both SNa winds and the radiation of the southern AGN. 
We discovered a nuclear, redshifted CO emission peaking in the mid point of the two nuclei, as it is the case for the CO emission at the systemic velocity. 
The large velocity dispersion, which reaches its maximum ($\sim$500 $\rm km~s^{-1}$)  in the mid point between the two nuclei, suggests that the gas might be highly turbulent in this region, although the presence of an unresolved rotation component cannot be ruled out.
}


\titlerunning{High resolution CO(1-0) mapping in \object{NGC6240}}
\authorrunning{Feruglio}
 \maketitle

\section{Introduction}

Galaxy encounters and AGN and starburst feedbacks represent key processes in the hierarchical galaxy formation
and evolution.  
Mergers modify the galaxy morphology, destabilize cold gas and
trigger both star formation and nuclear accretion onto
super-massive black holes (SMBHs), therefore inducing active galactic nuclei
(AGN) activity (Sanders et al. 1988, Barnes \& Hernquist 1996,
Cavaliere\& Vittorini 2000, Di Matteo et al. 2005). 
Simulations of major mergers of disk galaxies with small gas-to-star content ratio are able to
broadly reproduce the properties of local early type galaxies (Barnes \& Hernquist 1996,
Mihos \& Hernquist 1996, Bournaud et al. 2004, Hoffman et al. 2010). 
Mergers of gas-rich disk galaxies may easily form massive clumps due
to internal instabilities, and lead to the formation of irregular spheroids with high local velocity
dispersion (typically 150-200 $\rm km~s^{-1}$, Bournaud et al. 2011).

AGN and starburst feedbacks are also expected to affect evolution of galaxies (Silk \& Rees 1999, King 2010 and references
therein), by heating the interstellar medium (ISM) through winds, 
outflows and shocks,  thus inhibiting further accretion onto SMBH and quenching star formation on nuclear 
and possibly larger scales in the galactic disk of the host.
Radiative feedback from a luminous AGN and/or mechanical feedback from AGN jets could explain the low gas content of local massive galaxies and the galaxy bimodal color distribution (Kauffmann et al. 2003, Croton et al. 2006, Menci et al. 2006, Faucher-Giguere \& Quataert 2012).  
Evidence is mounting for AGN feedback as a mechanism contributing to the transformation of AGN host galaxies.
Massive, spatially extended outflows  have been recently discovered in several local (U)LIRGs and QSOs 
(Feruglio et al. 2010, Fisher et al 2010, Alatalo et al. 2011, Sturm et al. 2011, Aalto et al. 2012, Cicone et al. 2012, Maiolino et al. 2012, Feruglio et al. 2013: F13 hereafter). 

NGC6240 offers the opportunity to study the intermediate phase of a
merger event, between the first encounter and the final coalescence
(see e.g. Sanders \& Mirabel 1996, Mihos \& Hernquist 1996).  NGC6240
is a massive object, probably resulting from the merger of two gas
rich spirals.  The remnants of the bulges of
the progenitor galaxies, separated by $\sim2$\arcsec,
along a position angle of 40 degrees, are located in the central
region of the system (Engel et al. 2010).  
Each of them hosts an AGN, with SMBH masses exceeding $10^8~ \rm M_{\odot}$ (Engel et
al. 2010).  Both the AGNs are highly obscured by a hydrogen column density
of $\rm N_H>10^{24}$ cm$^{-2}$ (Compton-thick, Komossa et al. 2003).  
The southern AGN has an intrinsic luminosity of L(2-10 keV)$>10^{44}$ erg s$^{-1}$ (Vignati et al. 1999).
Observations of CO(1-0) , CO(2-1) and CO(3-2)
emission lines have shown that, unlike the stellar component, most of the molecular gas  is located in the region in
between the two nuclei (Tacconi et al. 1999, Iono et al. 2007, Engel et al. 2010, Feruglio et al. 2013). 
The origin of this  central concentration is still debated and fails to be reproduced by current simulations of major mergers of gas-rich galaxies.  
Tacconi et al. (1999) found evidence of coherent rotation in the central concentration and
ascribed it to a rotating molecular disk, although with unusually large velocity dispersion.    
Recent high resolution observations of CO(3-2) with the SMA resolved the central
concentration into two asymmetric peaks, separated by less than 1 arcsec, that
are interpreted as the two gas reservoirs of the merging galaxies, stripped off from the stellar systems, and in the phase of  tidally falling into the dynamical center  of the system (U et al. 2011).

NGC6240 displays the brightest line emission from hot molecular hydrogen ($\rm H_2$) among all LIRGs. 
The $\rm H_2$ emission peaks close to the southern AGN, unlike the CO emission, which reaches its maximum in the  mid point between the twoAGNs
(Tecza et al. 2000, Ohyama et al. 2003, Engel et al. 2010). 
The huge luminosity of $\rm H_2$  ($\rm 2 \times 10^{9}~L_{\odot}$, Egami et al. 1999, Tecza et al. 2000) is due to two main processes: 
a) the expanding motion of a shell-like structure around the southern nucleus (super-wind), and b) cloud-crushing at the interface of the two merging nuclei, where the super-wind interacts with the central concentration of molecular gas (Ohyama et al. 2000, 2003).

The large scale,  butterfly-shaped emission-line nebula seen in HST H$\alpha$ images
 is interpreted as evidence of a super-wind shock-heating the ambient ISM (Gerssen et al. 2004).  
The H$\alpha$ emitting filaments and
bubbles appear to trace a bipolar outflow, aligned east-westward,
extending up to 15-20 arcsec (7-10 kpc) from the nuclear region,
approximately perpendicular to the line connecting the two nuclei.
The superwind is probably powered by both the nuclear star-formation
and by the southern AGN.  F13 recently reports the detection of large scale
structures of CO, extended on 10 kpc scales, possibly tracing an outflow.  
Recently, Sturm et al. (private communication, in preparation) detected a P-Cygni profile absorption line of OH with  \emph{Herschel}-PACS, with velocities similar to those found in CO(1-0), unambiguously tracing a molecular outflow, although not spatially resolved due to the \emph{Herschel} PSF.

We present in this work high spatial resolution observations of \co~
in NGC6240 obtained with the IRAM Plateau de Bure Interferometer
(PdBI) in the A array configuration. F13  presents an analysis of the extended CO emission of 
NGC6240 on scales of $\sim$10 kpc around the nuclei, whereas in this work we focus on the nuclear region, close around the two AGN and  the two merging bulges.  All positions, fluxes and sizes given in the text are derived from fitting the visibilities in the uv plane, unless differently stated. 
A $\Lambda$CDM cosmology ($H_0=71$ km s$^{-1}$
Mpc$^{-1}$; $\Omega_M=$0.3; $\Omega_{\Lambda}=0.7$) is adopted.

\begin{figure}
\centering
\includegraphics[width=8cm]{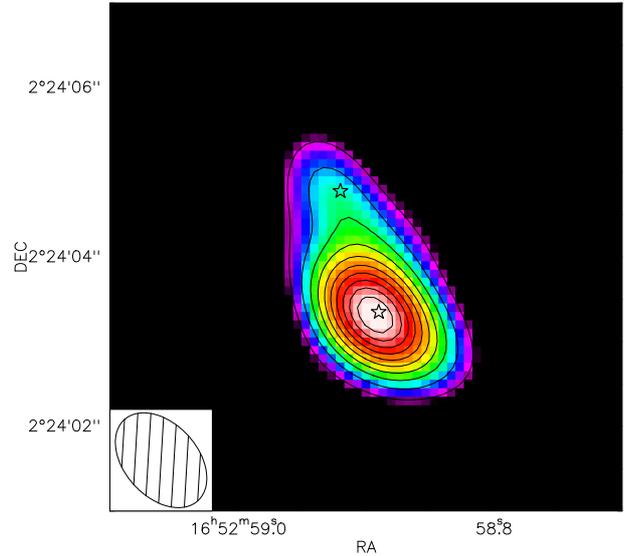}
\caption{3 mm continuum map of NGC6240 based on natural weighting.  The VLBI positions of the two
  AGN are indicated by stars. The contours are spaced by 3$\sigma$,
  starting from $5\sigma$. The oval shows the syhthesized beam .}
\label{cont}
\end{figure}

\section{PdBI observations and data analysis}

We observed with the PdBI the \co~ transition, redshifted to 112.516
GHz assuming a systemic velocity of 7339 $\rm km~s^{-1}$ (Iono et al. 2007),
corresponding to a redshift of $\rm z = $0.02448.  The observations were
carried out in January 2012 with 5 antennas in the extended (A) array
configuration.

Data reduction was performed using GILDAS.
The system temperatures during the observations were in the range
between 150 and 300 K.  The on-source time, after merging all the data
and after flagging of the bad visibilities, is 10 hours4.  The
absolute flux calibration relies on the strong quasars 3C273 and 3C279
and its accuracy is expected to be of the order 10\%.  The synthesized
beam, obtained by using natural weighting, is
$1.27\arcsec\times0.85\arcsec$, with PA$=41$ deg.  The achieved noise
level is 2 mJy/beam over 10 MHz (i.e. $\sim$26 $\rm km~s^{-1}$).  Robust
weighting yields a slightly higher spatial resolution (synthesized
beam of $1.20\arcsec \times 0.85\arcsec$), but degrades the
sensitivity to 2.5 mJy/beam in 10 MHz channels.  
The sensitivity of data from the D configuration alone is degraded by a factor  $\sim 1.5$ compared to the merged A+D configuration data. The gain in spatial resolution is, however, a factor 1.8 on average. 
This work makes use of the A configuration data alone, since a) we aim at reaching the maximum spatial resolution in the nuclear region of NGC6240, and b) the  CO emitting components presented in this paper are all detected with a significance $>10\sigma$ (see Section 3), so the  degraded sensitivity is not an issue.

\begin{figure*}
\centering
\includegraphics[width=\textwidth]{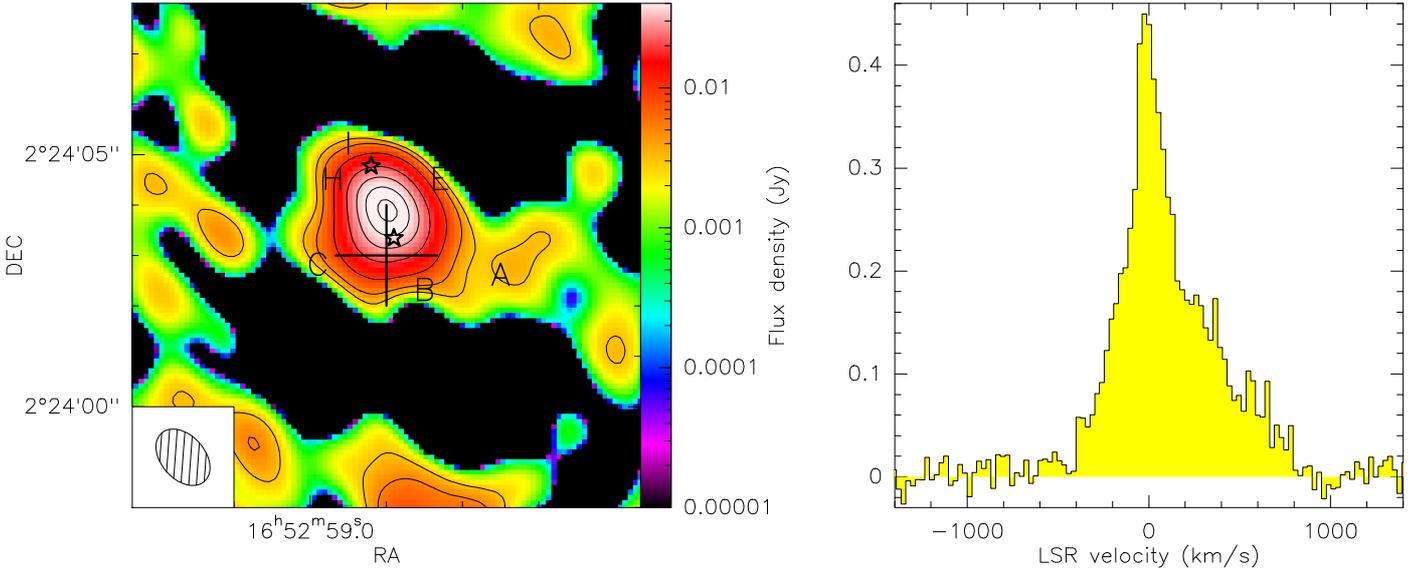}
\caption{Left panel: integrated \co~ map \object{NGC6240} (based on natural weighting, logarithmic scale in Jy).  
Stars indicate the positions of the two AGN.  The large cross indicate the phase tracking center. The contours are 3$\sigma$, 5$\sigma$, 10$\sigma$, 20$\sigma$, 30$\sigma$, 40$\sigma$, 50$\sigma$. Labels indicate the positions where
  we extracted the spectra shown in Figure 2.  Right panel:  continuum-subtracted CO line spectrum, integrated in a circular region of radius 3" centered on the peak of CO emission. The spectral channels are 26 $\rm km~s^{-1}$ wide.}
\label{spettro}
\end{figure*}

\section{Results}

\subsection{3 mm continua}
The 3 mm continuum was estimated in a region free from emission lines
(i.e. in the velocity range -3500 to -2000 $\rm km~s^{-1}$ and 2000 to 4000
$\rm km~s^{-1}$).  The 3 mm continuum is extended and resolved into two
components, each coincident with the VLBI position of the two AGN (Hagiwara et al. 2011, Figure
\ref{cont}). Most of the continuum emission is
  centered on the southern AGN. The strength of the 3 mm continuum, estimated by fitting two unresolved components is, 8.7$\pm0.9$ mJy for the southern component, and 2.9$\pm0.4$ mJy for the northern one.  This agrees within the expected accuracy of our measurements (10\%)
with the 3 mm continuum measured from the D-array configuration data ($12.7\pm1.3$ mJy, F13). 
This means that we expect little flux loss in the nuclear region due to the enhanced spatial resolution of the observations used here.
The continuum emission was subtracted from the total visibilities to obtain the line emission
spectra .
The radio spectral indexes (in the range 0.6-0.7)  based on 8 GHz (Colbert et al. 1994), 
 1 mm (Engel et al. 2010) and our 3 mm measurements support the synchrotron origin of the continua.

\begin{figure*}
\begin{center}
\includegraphics[width=\textwidth]{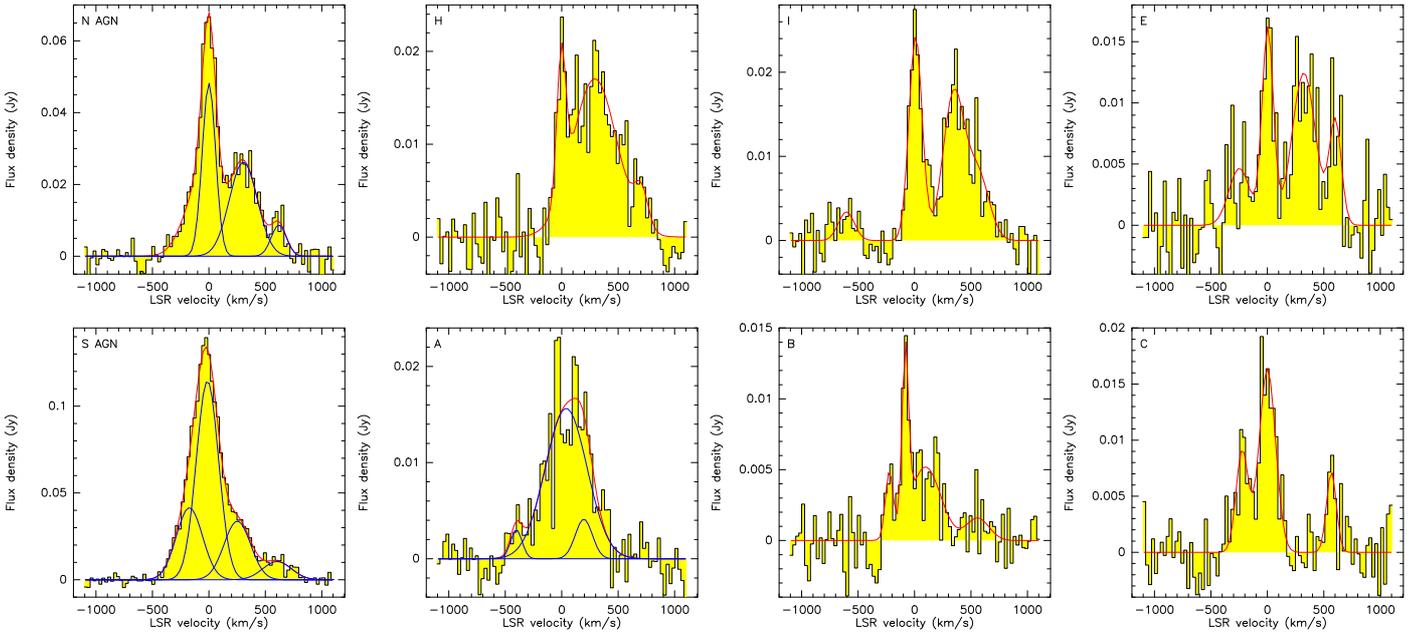}
\caption{Spectra at the positions marked in Figure1. N AGN, S AGN, A, B,
  C and E are the same positions as in Tacconi et al. (1999)
  maps. Gaussian fitting of the profile is shown in red (total fit). 
 Blue lines show the single gaussian components used in the fit for the panels where the total fit is not 
 self-explanatory.}
\label{abcde}
\end{center}
\end{figure*}

\begin{figure*}
\includegraphics[width=\textwidth]{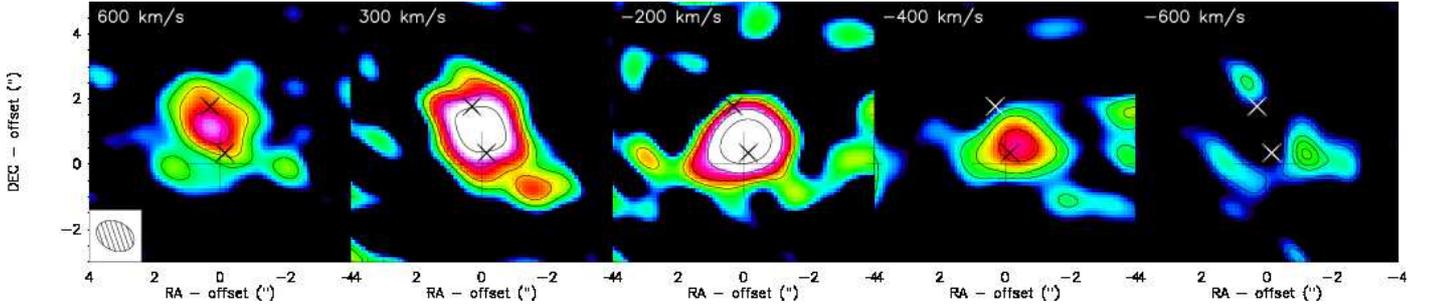}
\caption{Channel maps (derived using robust weighting) centered on the velocity peaks seen in Figure 3,
  and averaged over a velocity range equal to the FWHM of the
  corresponding gaussian fit.  Contours correspond to 3, 5, 10, 20$\sigma$ (3, 5, 6, 7$\sigma$ in the rightmost panel). The small crosses show the positions of the AGN. The large cross indicate the phase tracking center. }
\label{line-peaks}
\end{figure*}

 \begin{figure}[b]
\includegraphics[width=9cm]{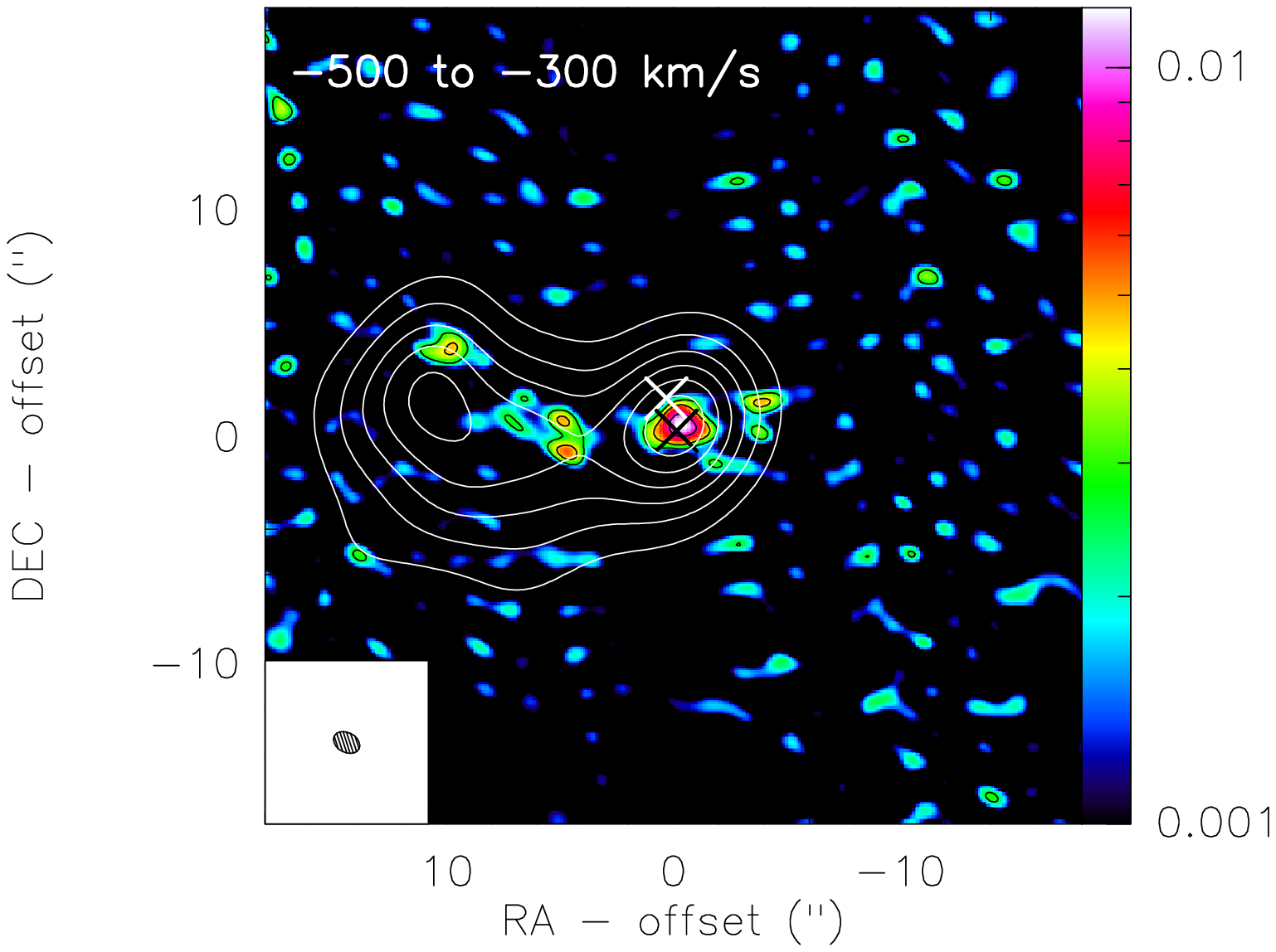}
\caption{Map of CO emission integrated over -500 to -300 km/s, and with size 36" squared.  Black contours 3, 5, 10, 20$\sigma$. The crosses show the positions of the two AGN. White contours (5 to 30$\sigma$, by 5$\sigma$) show the detection of the large-scale structures in the D configuration data, integrated in the range from -400 to -340 km/s (from Feruglio et al. 2o13).}
\label{large-structures}
\end{figure}

\subsection{CO(1-0) emission line}
Figure \ref{spettro}, left panel,  shows the continuum-subtracted, integrated map of CO(1-0).
The CO(1-0) spectrum, integrated over a circular area of radius 3" centered on the emission peak, is shown in the right panel. 
The positions of the two AGN nuclei from VLBI observations
(Hagiwara et al. 2011) are indicated by stars. 
The peak of the CO emission is centered in between the two AGN, closer to the southern AGN. 
Extended fainter
structures are visible on scales of 5$\arcsec$ along a position angle of 30 degrees, with similar morphology to those seen in the
CO(2-1) maps (Tacconi et al. 1999).  The CO  emission line shows a
complex profile, extending from $-500$ to 800 $\rm km~s^{-1}$ with respect to the
systemic velocity (F13).  
The integrated strength of the total CO(1-0) emission, by fitting the visibilities with three elliptical Gaussian components, is 321$\pm19$ Jy $\rm km~s^{-1}$ over $\rm FWZI=1400$ $\rm km~s^{-1}$. This is
in agreement with previous interferometric measurements (Bryant \& Scoville 1996, F13).

The spectra extracted from the regions labeled as A, B, C, E, H, I and
with stars are shown in Figure \ref{abcde}.  These include the positions
shown in Tacconi et al. (1999).  The spectra show similar line
profiles as those of CO(2-1), but extending out to larger velocities (out to 800 $\rm km~s^{-1}$)
on the red side of the line.  Several kinematic components are detected at different velocities. 
The spectrum at the position of the N AGN
(N AGN) shows a pronounced and well defined triple peak profile
(peaking at $\rm v\sim0, 300, 600$ $\rm km~s^{-1}$, Table 1).  The spectra at the positions E, H and I,
close to the N AGN, also show at least three spectral components, peaking at similar velocities to N AGN.
The spectrum at the position of the southern AGN (S AGN) shows a profile
similar to N AGN, but with a larger intensity ratio between the systemic and the $\rm v=300$ $\rm km~s^{-1}$ components. 
Conversely, the spectrum at the position A lacks the $\rm v\sim600$
component, which is present in the spectra of B and C. 
Blueshifted components at $\rm v\sim-200$ and $\rm v\sim-400$ appear at the
positions S AGN, A, B, C.

Figure \ref{line-peaks} shows channel maps  integrated over a bandwidth corresponding to the FWHM  
and centered on the velocities corresponding to the main peaks fitted in Table 1.
Both the systemic and the redshifted gas peak in the midpoint between the two nuclei , similarly to that found by 
Tacconi et al. (1999), Iono et al. (2007), and Engel et al. (2010) for CO(2-1) and CO(3-2) emissions. 
The $\rm v\sim300$ and $\rm v\sim600$ redshifted gas is elongated along a position angle of 40 degrees, 
with a size of $\sim5$ arcsec ($\sim2$ kpc). 
The integrated flux density  of the
high-velocity, redshifted component is 7.6$\pm0.5$ Jy $\rm km~s^{-1}$ in the velocity range  500 to 800 
$\rm km~s^{-1}$. (i.e. 2\% of the total CO emission).
 The blueshifted gas is instead found closer to the position of S AGN, peaking at -0.3, 0.6 arcsec offset from the phase tracking center. 
The integrated flux density of the blueshifted component, derived from uv-fitting of two Gaussian models, one elliptical and one circular, is $\rm 17.8\pm0.6~ Jy~ km/s$ over 300 km/s (i.e. in the velocity range $-200$ to $-500$ km/s), and its projected size is  $\rm \sim 2.65\pm0.2$ arcsec, corresponding to 1.3 kpc, in approximately the east-west direction, by combining the sizes of the two fitted components.
Figure \ref{large-structures} shows the same map as Figure 4 (panel corresponding to -400 km/s) but covering the entire field of view, in order to show the extended, blueshifted CO emitting structures discussed in F13. 
These extended structures  are recovered by the observations in A configuration, where the brightest clumps of emission are detected with significance $>5\sigma$ out to distances of $\sim 10$ arcsec eastwards and northeastwards of the nuclei, in a sort of filamentary pattern that develops in the same direction of the elongation of the compact component seen close to the S AGN. Only the more compact clumps are visible, the diffuse emission being filtered out by the interferometer. F13 showed that indeed the extended and compact diffuse gas components are connected. 
The geometrical and spectral matches suggest that the gas on large scales originates from the compact regions, although a projection effect cannot be a priori excluded.
F13 interpreted the extended structures as an outflow, originating from the S AGN, although the possibility that they are a tidal streamer of gas, left behind in the merging process, could not be ruled out.

\begin{figure*}
\includegraphics[width=\textwidth]{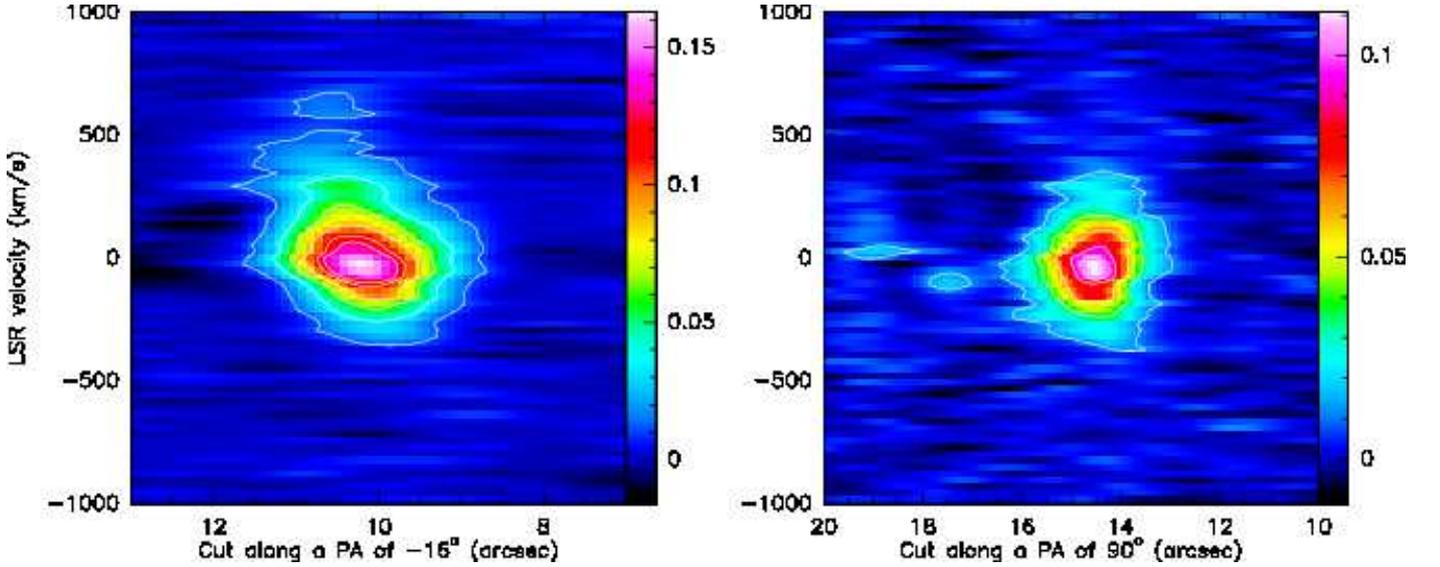}
\caption{Position-velocity plot cut along a PA of  -16 deg, i.e. the direction of the line connecting the 2 nuclei, from south-west to north-east (left panel), and along a PA of  90 deg  from north-west to south-east (right panel). Contours are 10$\sigma$ each, starting from 5$\sigma$ ($\sigma=2.5$ mJy/10MHz).}
\label{pv}
\end{figure*}

Figure \ref{pv} (left panel) shows the position-velocity cut (PV diagram) along a position angle $\rm PA =16$ degrees, i.e. along the line connecting the two nuclei, from south-west to north-east.   The gas with velocity larger than 500 km/s is shifted to the northeastern part of the system. Although no resolved rotation pattern is seen, the position shift of the gas in the range  $\pm300$ km/s along the cut, slightly larger than the beam,  suggests that there might be unresolved rotation in the central concentration of CO. 
The right panel of Figue \ref{pv} shows a cut along  a $\rm PA=90$ degrees,  corresponding to the direction where the blueshifted emission is elongated,  and intercepting the position of the S AGN.  Here as well, we do not find evidence for a coherent rotation. We find, however, that the gas with velocity between -200 and -400 km/s is seen everywhere along the cut,  on scales of $\sim2.5$ arcsec, as found also by fitting of the visibilities. A separated component (5$\sigma$) with velocity $\sim -200$ km/s is seen at 3 arcsec eastward of the nuclei.

\begin{figure*}[t]
\centering
\includegraphics[width=\textwidth]{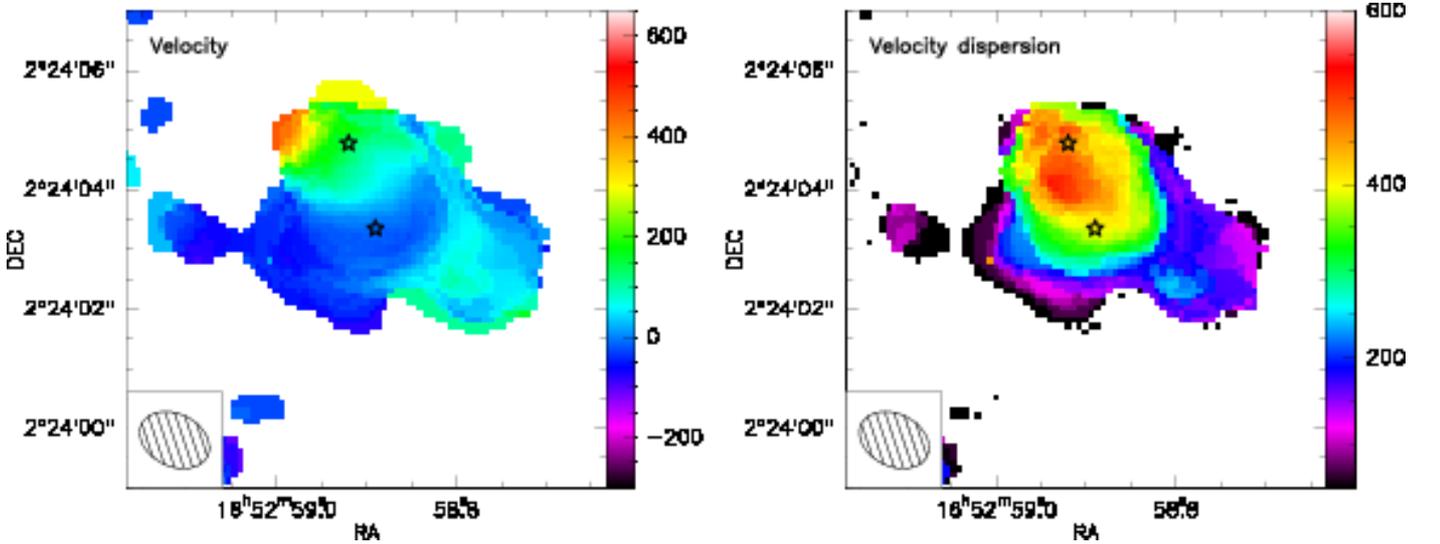}
\caption{Left: CO(1-0) velocity distribution (moment 1) map of NGC6240. The units are in $\rm km~s^{-1}$.  
Right:  velocity dispersion (moment 2) distribution map. Both derived by using robust weighting to enhance the spatial resolution.
Symbols as in Figure 1.}
\label{width}
\end{figure*}

The offset in position between the peak of the redshifted and blueshifted emission 
is shown in Figure \ref{width} (left panel), which shows the velocity  (moment 1) distribution map of CO(1-0).  
The velocity gradient between the two AGN is $\sim 150$ km/s.
Figure \ref{width} (right panel) shows the line-of-sight velocity dispersion (moment 2) map. 
This reaches its maximum ($\sim$500 $\rm km~s^{-1}$) in the central CO concentration, and in the region around the N AGN.

\begin{table}[b!]
\centering
\caption{\label{t1}Parameters of the Gaussian fit of the CO-emitting components shown in Figure 3.}
\tiny
\begin{tabular}{lccc}
\hline
Component   &  $\rm I_{peak}$  &   v, CO(1-0)   &    FWHM \\
              & [Jy]   &  [$\rm km~s^{-1}$]  & [$\rm km~s^{-1}$]  \\
\hline
N AGN  &    4.8$\pm2~10^{-2}$    &   $0.\pm5.0$              &    227$\pm145$ \\
    &                2.6$\pm0.2~10^{-2}$   &   305$\pm49$      &    470$\pm179$ \\
    &                8.7$\pm3~10^{-3}$    &   619$\pm40$       &   260$\pm106$  \\
\hline
E &    4.6$\pm2~10^{-3}$   &  $-248\pm50$     &     $355\pm167$ \\
    &   1.6$\pm0.2~10^{-2}$   &   $2.8\pm9.5$       &   $193\pm57$ \\
    &   1.2$\pm0.2~10^{-2}$   &  $ 322\pm18$       &   $400\pm105$ \\
    &   8.4$\pm3~10^{-3}$   &  $602\pm35$       &   $204\pm79$\\
\hline
H &    1.5$\pm0.3~10^{-2}$    &   $-4.0\pm7.0$  &      $152\pm50$\\
    &    1.7$\pm0.1~10^{-2}$     &   $292\pm21$     &      $755\pm171$  \\
    &    4.0$\pm3~10^{-3}$    &  $705\pm67$      &     $234\pm125$\\
\hline
I  &    2.4$\pm0.2~10^{-2}$  &   $8.6\pm4.6$    &     $226\pm41$ \\
   &    1.6$\pm1~10^{-2}$   &   $338\pm28$    &      $347\pm74$ \\
   &    8.3$\pm3~10^{-3}$   &   $540\pm136$   &      $ 441\pm274$ \\
   &   $3.4\pm2~10^{-3}$   &   $-603\pm65$   &      $272\pm297$ \\
\hline
S AGN     &  4.2$\pm7~10^{-2}$   &  $-173\pm408$    &     $ 444\pm393$   \\
      &  $0.1\pm0.1$           &  $-13\pm59$       &    $379\pm146$        \\
      & $3.4\pm0.5~10^{-2}$   &  $251\pm69$     &     $450\pm224$     \\
      &  1.0$\pm0.3~10^{-2}$   &  $604\pm112$     &      $488 \pm271$    \\
\hline
A   &    3.0$\pm3~10^{-2}$   &  $-402\pm74$    &     $194\pm128$    \\
   &    1.6$\pm0.3~10^{-2}$   &   $39\pm42$        &   $711\pm164$     \\
   &    4.1$\pm5~10^{-3}$    &  $195\pm85$      &    $289\pm172$     \\
\hline
B   &    4.6$\pm2~10^{-3}$   &  $-224\pm23$     &     $129\pm61$     \\
   &    1.2$\pm0.2~10^{-2}$    &  $-82\pm4$       &   $108\pm43$        \\
   &   5.2$\pm1~10^{-3}$   &    $96\pm54$       &  $521\pm220$        \\
    &   1.6$\pm1~10^{-3}$     & $559\pm103$        &  $384\pm224$         \\
\hline
C   &    9.1$\pm2~10^{-3}$   &  $-224\pm19$   &       $207\pm77$     \\
      &   1.6$\pm0.1~10^{-2}$    &    $-4\pm8$        &  $278\pm61$         \\
      &   7.2$\pm2~10^{-3}$    &  $567\pm28$       &  $ 167\pm56$        \\
\hline
\end{tabular}
\end{table}


\section{Discussion}

We have mapped the CO(1-0) emission line in NGC6240 with the extended
A configuration of the PdBI, reaching an average spatial resolution of
$\sim 500$ pc at the redshift of the source.  The 3 mm continuum shows
two components, coincident with the position of the two AGN.
Our 3 mm measurements confirm the non thermal synchrotron
origin of the continua (Engel et al. 2010 and references therein). 

Thanks to the large bandwidth available today at the PdBI (3.6 GHz), we have
detected a broad CO(1-0), reaching larger redshifted velocities
than those previously reported (Tacconi et al. 1999, Iono et al. 2007, Engel et al. 2010).  
The CO emission, at different locations of the system, show similar line profiles to 
those of CO(2-1) (Tacconi et al. 1999, Engel et al. 2010), but with larger line width, and exhibit several kinematic components (blue and redshifted, compared to the systemic one, Table 1). 
We confirm that the molecular gas is concentrated in between the two AGN, in the so-called central concentration.  

The total integrated luminosity of the CO line is $\rm L'(CO) = 8.6\times 10^{9}~ K~ km~ s^{-1}~
pc^2$, including the central concentration and the extended structure.  
Assuming the conversion factor tipically used for ULIRGs, $\alpha=0.8$ (Solomon et al. 1997),
we derive a total mass of molecular gas of  $\rm M(H_2)=6.8\pm0.7\times 10^{9}~M_{\odot}$. 
This is in agreement with the estimate reported by Iono et al. (2007), $\rm 6.8\pm1.7\times 10^{9}~M_{\odot}$, based on CO(3-2).
The luminosity of CO in the central 1 kpc region is 
$\rm L'(CO) = 5\times 10^{9}~ K~ km~ s^{-1}~ pc^2$ over a line width of 
1400 $\rm km~s^{-1}$.  This converts into a molecular gas mass of
$\rm M(H_2)=4\times10^{9}~M_{\odot}$, consistent with the mass derived from CO(2-1) (Engel et al. 2010).

We report the discovery of a new nuclear component of CO(1-0) at redshifted velocities
larger than 500 $\rm km~s^{-1}$ and up to 800 $\rm km~s^{-1}$ with respect to the systemic CO-derived velocity.  
The luminosity of this component is 
$\rm L'(CO)=2\times 10^{8}~ K~ km~ s^{-1}~ pc^2$.  The inferred mass of
molecular gas is $\rm M(H_2)=1.6\times 10^{8}~M_{\odot}$, assuming $\alpha=0.8$.  
The possible existence of this component was suggested by Tacconi et al. (1999) and Engel et al. (2010) based on CO(2-1), 
although no direct evidence of such emission has been reported due to the small bandwidth.  
Interestingly, the redshifted emission peaks in the
region between the two AGN, northwards compared to the
blueshifted gas, which is concentrated closer to the S AGN (Figure 4), similarly to CO(2-1).  
This fact was interpreted as a evidence of a rotating, turbulent molecular disk (Tacconi et al. 1999). 
In contrast, Engel at al. (2010) argued that it was more likely to be a bridge of gas, which would be prominent for a merger geometry
that is not too far from co-planar and prograde. They also noted, however, that for the gas mass to be consistent with models, the CO-to-$\rm H_2$conversion factor would have to be high, as it is the case of non-virialised clouds.
Our data can hardly be reconciled with one turbulent, rotating disk.
The large velocity observed (800 $\rm km~s^{-1}$), and the  lack of a corresponding blueshifted component, rule out the rotating disk hypothesis. 
U et al. (2011) suggest that the large velocity gradients seen in the nuclear region are two rotating disks (the original gas reservoirs of the merging bulges) orbiting the dynamical center of the system. Limited by the spatial resolution of our data, we cannot confirm or discard this hypothesis.
Current hydrodynamical simulations of mergers of gas rich galaxies  (Bournaud et al. 2011) predict the formation of clumps, producing an emission line profile with multiple components. The velocity dispersions due to turbulence can reach in these systems 150-200
$\rm km~s^{-1}$, and even 500 $\rm km~s^{-1}$, when projected along some lines of sight.
These simulations describe mergers of high-redshift galaxies, like SMGs, for which the molecular gas is expected to be more turbulent than local galaxies. 
NGC6240 is a merger of two massive, gas-rich galaxies, a rare case in the local universe, but common among high redshift galaxies.
The origin of the huge concentration of gas in the mid point of the two nuclei is, however, still debated, and not explained by current models of merging systems.

The blueshifted emission with velocity $-200$ $\rm km~s^{-1}$ peaks at a position $0.4$ arcsec
north of the S AGN, at the same position of the $\rm H_2$ peak (Ohyama et al. 2000, 2003).
The blueshifted CO with $\rm v=-400$ $\rm km~s^{-1}$ is also found close to the S AGN, matching the $\rm H_2$ outflowing component at similar velocity (Ohyama et al. 2000, 2003, Engel et al. 2010).  
The $\rm H_2$ blueshifted components with velocities -200 and -500 $\rm km~s^{-1}$ are interpreted as arising from super-wind activity in the southern nucleus. 
These spatial and kinematic matches strongly suggest that the CO blueshifted emission is associated with the H$_2$ emitting expanding shell, originating from the southern nucleus.
The blueshifted component has $\rm L'(CO)=4.7 \pm 0.2 \times 10^{8}~K~ km~ s^{-1}~ pc^2$, corresponding to a mass of molecular gas of 
$\rm M(H_2)=2.4\pm 0.2 \times 10^{8}~M_{\odot}$, assuming $\alpha=0.5$ (Weiss et al. 2001, Feruglio et al. 2010). 
Assuming that the outflow is symmetric along the major axis (bi-conical geometry), $\rm R_{of}=0.6\pm0.1$ kpc, and the terminal de-projected velocity equal to the maximum blueshifted velocity $\rm v_{of} = -400~ km/s$, we derive a lower limit mass loss rate using the relation of Feruglio et al. (2010) and Maiolino et al. (2012):

\begin{equation}
\rm \dot M = \frac {3 \times v_{of} \times M(H_2) }{R_{of}} \simeq 500~ M_{\odot} yr^{-1} 
 \end{equation}

\noindent
assuming that the (bi)cone is uniformly filled with gas (or with gas clouds), and that there is no mass loss through the lateral sides of the cone.  
The kinetic power entrained in the outflow is $\rm P_{k,OF} = 0.5 \times \dot M \times v_{of} \eqsim 3\times 10^{43}~ erg/s$ . 
The nuclear star-formation rate measured from FIR SED fitting is $60\pm30~ \rm M_{\odot}~yr^{-1}$ (Yun \& Carilli 2002). 
The derived mass loading factor of the molecular gas, $\rm \dot M/SFR $ in the range 8-17, suggests that, in addition to the SNa winds, the AGN contributes to the molecular outflow. 
In fact, according to the relation of Veilleux et al. (2011), at the rate found for NGC6240,  SNe can power winds with a maximum kinetic power of $\rm P_{k,SF} = 4-7 \times 10^{42}~ erg/s$.
By fitting  in the same spectral range, the nuclear component from the D configuration data, and using the same assumptions, we get a mass of molecular gas $\rm M(H_2)=  1.8\pm0.4 \times 10^{8}~M_{\odot}$, and a mass loss rate in the range 300 to 450 $M_{\odot} yr^{-1}$. Accounting for the $10\%$ uncertainty in the  flux calibration the two estimates are consistent. 
Conversely, we find a discrepancy when comparing the compact outflow with that extended eastwards on 10 kpc scale (F13), which has a $\rm \dot M \sim 100~M_{\odot} yr^{-1}$.  We discuss here possible reasons for this discrepancy, in addition to the mis-knowledge of the conversion factor $\alpha$, which might be different in the compact region compared to the extended one. Contrary to what assumed in our simple scenario, the eastern extended cone does not appear to be uniformly filled with gas, when seen with high spatial resolution: it shows clumps of emission of different sizes and brightnesses. Moreover, if significant amount of gas is lost through the lateral sides of the cone, the mass loss rate is not conserved with increasing radius.

Finally, the similar spatial resolution of our maps and of those of CO(2-1)  
(Tacconi et al. 1999)  allows us to compare the line strengths in different locations of the system, i.e. the central kpc and 
the extended region on scales of 3 arcsec around the nuclei. 
Figure 1 shows that the CO(1-0) emission exhibits a very similar extension and structures to the CO(2-1) one.  
We measure the total emission in two apertures centered on the emission peak, with radius 1 and 1.5 arsec each. 
In the central 1 kpc region around the central concentration we find
$\rm L'(CO) =3.4\times 10^{9}~ K~ km~ s^{-1}~ pc^2$ in the velocity range -500 to 500 $\rm km~s^{-1}$.  
 Tacconi et al. 1999) measured a flux of $\rm CO(2-1) =528~ Jy~ km~s^{-1}$  for the central 1 kpc region, and 
 a total flux density of 1220 Jy $\rm km~s^{-1}$ in a 1.5 kpc region, in the same velocity range.
The ratio $\rm L~ CO_{(2-1)} / L ~CO_{(1-0)}= 1.0$ for the central concentration, indicating that there the gas is moderately optically thick. 
The integrated flux density in the extended region around the two nuclei is $\rm L'(CO)=1.2\times 10^{9}~ K~ km~ s^{-1}~ pc^2$, which gives $\rm L~ CO_{(2-1)} / L~ CO_{(1-0)}\sim 3.8$, suggesting that the gas here is instead optically thin.

\section{Summary}
We present high spatial resolution CO(1-0) imaging of the luminous infrared galaxy NGC6240 
obtained with the PdBI. We provide our main findings below: 
\begin{enumerate}

\item  We find a broad CO(1-0) line profile, with a maximum velocity 800 $\rm km~s^{-1}$ and $\rm FWZI=1400$ $\rm km~s^{-1}$, with several kinematic components, witnessing the complexity of the gas dynamics in this source. The CO emission peaks in the midpoint between the two nuclei, as found in previous studies (Engel et al. 2010 and references therein).
This component is not predicted by recent hydrodynamical simulations of mergers between gas rich galaxies (Bournaud et al. 2011).

\item The blueshifted CO emission with -200 and -500 $\rm km~s^{-1}$, peaks close to the S AGN, at the same position where a $\rm H_2$ outflow is found. 
Based on this, and on the fact that an outflow is also seen, with similar velocities, in absorption with \emph{Herschel}-PACS (Sturm et al. in prep.), we regard this result as a clear indication of a massive molecular outflow, expanding from the southern nucleus.
Its mass loss rate is $\sim 500~ \rm M_\odot yr^{-1}$, and the mass loading factor of the molecular gas, $\rm \dot M/SFR$, is of the order 10. 
The CO outflow is likely connected to the $\rm H_2$ superwind around the southern nucleus (Ohyama et ao 2000, 2003), and to the large scale CO outflow, found on scales of 10 kpc, showing similar velocities (F13) and a filamentary structure that appears connected to the S AGN.  The molecular compact outflow is likely driven by both AGN and SNe.

\item The redshifted CO emission, which shows a maximum velocity of 800 km/s,  peaks in the mid point of the two nuclei, as it is the case for the systemic CO. The most redshifted gas is shifted towards the northern part of the system (close to N AGN).
The gas is probably turbulent in the nuclear region, as supported by the large velocity dispersion, which reaches its maximum ($\sim$500 $\rm km~s^{-1}$)  in the central CO concentration. 
We suggest that this component of the molecular gas is flowing towards, or perhaps orbiting the center of mass of the system, where probably the two gas reservoirs will eventually merge.
 Limited by the spatial resolution of our data, however, we cannot conclude about the presence or not of rotation in the centre of the system.

\end{enumerate}

Interferometric observations with higher spatial resolution with ALMA will certainly add crucial constraints on the gas dynamics in NGC6240.

\begin{acknowledgements}
We thank Dennis Downes for useful discussions. F.F. acknowledges support from PRIN-INAF 2011.
\end{acknowledgements}


\begin{thebibliography}{} 

 \bibitem {} Aalto, S., Garcia-Burillo, S., Muller, S., Winters, J. M., van der Werf, P.  et al.  2012, A\&A 537, 44 
 \bibitem {}   Alatalo, K. A., Davis, T. A., Young, L. M., Heiles, C., Blitz, L.  et al. 2011, ApJ, 735, 88
  \bibitem {} Barnes, J. E. \& Hernquist, L.  1996, ApJ 471, 115
    \bibitem {} Bournaud, F., Combes, F. \& Jog, C. J. 2004, A\&A, 418, L27
 \bibitem {} Bournaud, F., Chapon, D., Teyssier, R., Powell, L. C., Elmegreen, B. G. et al. 2011, ApJ, 730:4
 \bibitem {} Bryant, P. M., \& Scoville, N. Z. 1996, ApJ, 457, 678
   \bibitem {}Cavaliere, A. \& Vittorini, V.  2000, ApJ 543, 599
  \bibitem {} Cicone, C., Feruglio, C., Maiolino, R., Fiore, F., Piconcelli, E., Menci, N.  et al.  2012, A\&A in press, arXiv1204.5881C
  \bibitem {} Colbert, E. J. M. , Wilson, A. S. \& Bland-Hawthorn, J. 1994, ApJ, 436, 89 
  \bibitem {} Croton, D. J., Springel, V., White, S. D. M., De Lucia, G., Frenk, C. S.   et al. 2006, MNRAS 367, 864 
 \bibitem{} Di Matteo, T., Springel, V. \& Hernquist, L. 2005, Nature, 433, 604
  \bibitem {} Engel, H., Davies, R. I., Genzel, R., Tacconi, L. J., Hicks, E. K. S.  et al.  2010, A\&A 524, 56 
 \bibitem {}  Faucher-Figuere, C-A. \& Quataert, E. 2012, MNRAS, 425, 605
 \bibitem {}  Feruglio, C., Maiolino, R., Piconcelli, E., Menci, N., Aussel, H. et al. 2010, A\&A 518, 155
 \bibitem {}  Feruglio, C., Fiore, F., Maiolino, R., Piconcelli, E., Aussel, H. et al. 2013, A\&A, 549A, 51
 \bibitem {} Fischer, J., Sturm, E., Gonzales-Alfonso, E., Gracia-Carpio, J., Hailey-Dunsheath, S.  et al., 2010,  A\&A, 
 \bibitem {} Gerssen, J., van der Marel, R. P., Axon, D., Mihos, J. C., Hernquist, L. et al.  2004, AJ, 127, 75
\bibitem {} Hagiwara, Y., Baan, W. A. \& Klockner, H. 2011, AJ 142, 17
 \bibitem {} Hoffman, L., Cox, T. J., Dutta, S., \& Hernquist, L. 2010, ApJ, 723, 818
 \bibitem {} Iono, D., Wilson, C. D., Takakuwa, S., Yun, M. S., Petitpas, G. R.  et al.  2007, ApJ, 659, 283 
  \bibitem {} Kauffmann, G., Heckman, T. M., Tremonti, C., Brinchmann, J., Charlot, S. et al. 2003, MNRAS, 346, 1055
\bibitem {}  King, A. R.   2010, MNRAS, 402, 1516 
\bibitem {} Komossa, S., Burwitz, V., Hasinger, G., Predehl, P., Kaastra, J. S. et al.  2003, ApJ 582, 15
 \bibitem {} Maiolino, R., Gallerani, S., Neri, R., Cicone, C., Ferrara, A. et al. 2012, MNRAS 425, 66
  \bibitem {} Menci, N., Fontana, A., Giallongo, E., Grazian, A. \& Salimbeni, S. et al. 2006, \apj, 647, 753
 \bibitem {} Menci, N., Fiore, F., Puccetti, S. \& Cavaliere, A. 2008, \apj, 686, 219 
  \bibitem {} Mihos, J. C.  \& Hernquist, L.  1996, ApJ 464, 641
  \bibitem {} Ohyama, Y., Yoshida, M., Takata, T., Imanishi, M., Usuda, T. et al. 2000, PASJ, 52 563
   \bibitem {} Ohyama, Y., Yoshida, M. \& Takata, T. 2003, AJ 126, 229
 \bibitem {} Riffel, R. A., Storchi-Bergmann, T., Winge, C., McGregor, P.J., Beck, T. et al. 2008, MNras, 385, 1129
  \bibitem {} Sanders, D. B., Soifer, B. T., Elias, J. H., Madore, B. F., Matthews, K. et al., 1988, \apj, 325, 74
 \bibitem {}  Sanders, D. B. \& Mirabel, I.F.  1996 ARA\&A, 34, 749
 \bibitem {} Scoville, N.  Z., Frayer, D. T., Schinnerer, E. \& Christopher, M. 2003, \apj, 585, L105 
 \bibitem {} Silk, J. \& Rees, M. J.  1998, A\&A, 331, L1 
  \bibitem {} Sturm, E., Gonzalez-Alfonso, E., Veilleux, S., Fischer, J., Gracia-Carpio, J.  et al. 2011, A\&A 518, 36
   \bibitem {} Solomon, P. M. et al. 1997, ApJ 478, 144 
 \bibitem {} Tacconi, L., Genzel, R., Tezca, M., \& Gallimore, J. F.  1999, ApJ 524, 732
 \bibitem {} Tecza, M., Genzel, R., Tacconi, L. J.,  Anders, S., Tacconi-Garman, L. E. et al. 2000, ApJ, 537, 178
 \bibitem {} Van der Werf, P., Genzel. R., Krabbe, A., Blietz, M, Lutz, D. et al. 1993, ApJ, 405, 522 
 \bibitem {} U, V. , Wang, Z., Sanders, D., Fazio, G., Chung, A. et al. 2011, ASPC, 446, 97
 \bibitem {} Vignati, P. et al.  1999, A\&A 349, 57
 \bibitem {}  Yun, M. S., \& Hibbard, J. E. 2001, in Gas and Galaxy Evolution, ed. J. E. Hibbard, M. Rupen, \& J. H. van Gorkom, ASP Conf. Proc., 240, 866 
 \bibitem {} Weiss, A., Neininger, N., H\'uttemeister, S., \& Klein, U. 2001, A\&A, 365, 571



  
  \end{thebibliography}
\end{document}